\documentclass[aps,amsmath,showpacs,amsfonts,10pt]{revtex4}
\usepackage{epsfig,graphicx}
\usepackage[english]{babel}
\usepackage{amsfonts}
\usepackage{amsmath}
\usepackage{latexsym}
\usepackage{graphics,bm}
\usepackage{natbib}
\usepackage{dcolumn}
\usepackage{bm}
\usepackage{rotating}
\begin{document}

\title{Degenerate Bose Gas: A New Tool for Accurate Frequency Measurement}

\author{A. B. Bhattacherjee$^{1}$ and S. De$^{2}$\footnote{Electronic mail: subhadeep@mail.nplindia.org}}

\address{$^{1}$School of Physical Sciences, Jawaharlal Nehru University, New Delhi-110067, India}\address{$^{2}$CSIR-National Physical Laboratory, Dr. K. S. Krishnan Marg, New Delhi - 110012, India.}

\begin{abstract}
We propose a new way of detecting frequencies using
superradiant Rayleigh scattering from degenerate Bose gases.
A measurement of the time evolution of population at the
initial momentum state could determine an unknown frequency with
respect to a known one at which the pump laser's frequency
modulates. A range of frequencies from kHz to $\sim$MHz could be
determined with a fractional uncertainty $10^{-6}$.
\end{abstract}


\maketitle

\begin{figure}
\begin{center}
\resizebox{0.4\textwidth}{!}{
\includegraphics{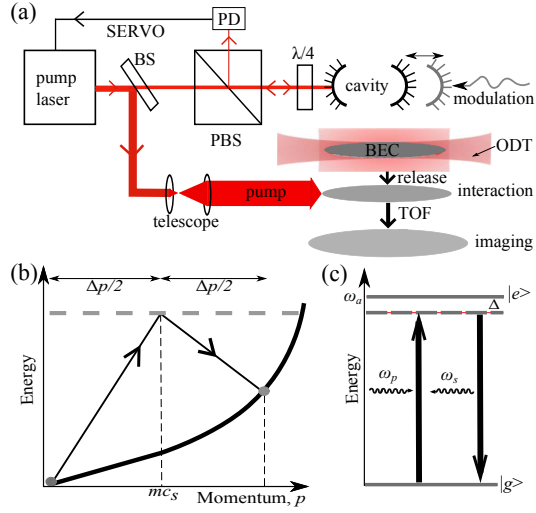}
}
\end{center}
\caption[Experiment]{In superradiant scattering: (a)
schematic of the experiment, (b) population transfer to the higher
momentum state and (c) interaction of a two level atom with a
far off-resonant photon, where $c_s$ is speed of phonons in the
BEC.} \label{Fig:Experiment}
\end{figure}

Bose Einstein condensate (BEC) atoms have long-range spatial coherence \citep{Bloch_Nature_2000} which offers the possibility to study quantum optics in a new domain where the atom-photon interactions could be altered \citep{Moore_PRA_1999}. The interest in superradiant Rayleigh scattering and matter-wave amplification originates due to long coherence time of the condensate \citep{Inouye_Science_1999,Kozuma_Science_1999,Schneble_Science_2003}. The first photon scattered from the condensate
leaves a perturbed BEC, which induce the next photons to scatter along
the same direction. Superradiant radiation occurs due to
spontaneous Rayleigh scattering of photons from an
elongated BEC \citep{Inouye_Science_1999}. Due to momentum
conservation atoms end up in a different momentum state after each
scattering \citep{Inouye_PRL_2000}. These scatterings are equally
probable in all directions so atoms could go to various
momentum states. However the elongated shape of the condensate
introduces self amplification of a particular mode and forms a
matter wave grating \citep{Inouye_Nature_1999,Kozuma_Science_1999}.
This redistribute atoms in different momentum states
\citep{Kozuma_Science_1999,Schneble_Science_2003}. Reverse
avalanche occurs in the scattered photon mode which is the well known Dicke
superradiance \citep{Dicke_PR_1954}. This mechanism is also known as
collective atomic recoil lasing (CARL). That was first described
for a free electron laser system
\citep{Bonifacio_NIMPRA_1994,Bonifacio_PRA_1994}, however it
was first observed in an atomic vapor cell
\citep{Lippi_PRL_1996}. Damping of photons
limit intensity of the superradiance
\citep{Slama_PRA_2007,Motsch_NJP_2010}.  This can be improved by
storing the BEC in an optical cavity which will enhance the atom-photon
interactions. The influence of atomic motion on the
superradiant light scattering from a moving BEC was investigated
theoretically and experimentally
\citep{Bonifacio_OC_2004,Fallani_PRA_2005}.
In this letter we propose a new way of detecting frequencies using
the technique of population transfer to a different momentum state in
superradiance. An analytic model is described for detection of an
unknown frequency with respect to a reference one. Also we give an
experimental scheme for realization of our proposal.
We consider the superradiant radiation occurs when a
off-resonant pump laser beam impinges along the axial direction of
a cigar-shaped BEC. The atom-photon interaction initiates
after the condensate is released from its confining potential in
an optical dipole trap (ODT), as shown in Fig.
\ref{Fig:Experiment} (a). An ensemble of degenerate atoms maintain
its elongated shape even after 1-10 ms time of flight (TOF).
Atom-photon interactions are fast processes than the time at
which superradiant dynamics develop. This is limited by the
decay of pump photons from the condensate and is faster
than the expansion rate of the BEC. We
consider BEC is very dilute after expansion and hence
intra-atomic interactions are ignored. In an anisotropic condensate,
correlation between scattered photons is enhanced when scattering
occurs along the elongated direction ($z$-axis) of the condensate.
This is known as end-fire mode in the regime of Dicke
superradiance \citep{Dicke_Conf_1964}. Here an atom scatters
photon from the pump laser beam of wave-vector $k_p$ and
propagates along the $z$-direction. The recoil photon along the
opposite direction results in a net momentum change $\Delta p = 2
\hbar k_p$ of the atom as shown in Fig. \ref{Fig:Experiment} (b).
Internal state of atoms do not change but they spread between two
momentum states separated by $\Delta p$. Hence atomic center of
mass motion changes but different momentum states get coupled.
The Hamiltonian describing the superradiant Rayleigh scattering is
    \begin{eqnarray}
    \label{eq:Hamiltonian}
    \hat{H} = \hat{H}_{a} + \hat{H}_{p} + \hat{H}_{a-p},
    \end{eqnarray}
which consists of energy associated with free atoms $\hat{H}_{a}$,
pump laser field $\hat{H}_{p}$ and interaction of atoms with the
photons in the pump laser light $\hat{H}_{a-p}$. We will explire
dynamics along the $z$-axis of the elongated
condensate. For a two level atom of mass $m$ the atomic
Hamiltonian is
    \begin{eqnarray}
    \label{eq:AtomicHamiltonian}
    \hat{H}_{a} &=& \int dz \bigg[ \hat\psi_{|g\rangle}^{\dagger}(z)
    \left ( - \frac{\hbar^{2}}{2m} \frac{\partial^{2}}{\partial z^{2}} \right )
    \hat\psi_{|g\rangle}(z) \nonumber \\
    &~&
    + \, \hat\psi_{|e\rangle}^{\dagger}(z)\left ( - \frac{\hbar^{2}}{2m }
    \frac{\partial^{2}}{\partial z^{2}}+\hbar \omega_{a} \right ) \hat\psi_{|e\rangle}(z)
    \bigg],
    \end{eqnarray}
where $\hat\psi_{|g\rangle}$ ($\hat\psi_{|e\rangle}$) and
$\hat\psi_{|g\rangle}^{\dagger}$
($\hat\psi_{|e\rangle}^{\dagger}$) are annihilation and creation
operators of the atom in its ground (excited) states which are
separated by a frequency $\omega_a$ and $\hbar$ is the Planck
constant. The bosonic field operators satisfy the equal time
commutation relations $[\hat\psi_{j}(z), \,
\hat\psi_{j'}^{\dagger}(z')]=\delta_{j,j'} \delta(z,z')$
also $[\hat\psi_{j}(z), \,
\hat\psi_{j'}(z')]=[\hat\psi_{j}^{\dagger}(z), \,
\hat\psi_{j'}^{\dagger}(z')]=0$, where subscripts $j, \, j'$
refers to both $|g\rangle$ and $|e\rangle$ states respectively. In
this case one can neglect the excited state population in
Eq. \eqref{eq:AtomicHamiltonian} and spontaneous emission since
frequency of the pump laser $\omega_p \ll \omega_a$. Hamiltonian of
the optical pump field of strength $\eta$ is
    \begin{eqnarray}
    \label{eq:PumpHamiltonian}
    \hat{H}_{p}=-i \hbar \eta (\hat{c}_p-\hat{c}_p^{\dagger}),
    \end{eqnarray}
where $\hat{c}_p$ ($\hat{c}_p^{\dagger}$) are the annihilation
(creation) operator of photons, which satisfy the commutation
relation $[\hat{c_p}, \, \hat{c_p}^{\dagger} ]=1$. The interaction
Hamiltonian of atoms with the photons in the pump laser light is
    \begin{eqnarray}
    \label{eq:AtomPhotonHamiltonian}
    \hat{H}_{a-p} = &-& i \hbar g c_p \bigg[ \int dz \, \hat{\psi}^{\dagger}_{|e\rangle}(z)
    e^{i k_p z} e^{i \phi(t)} \hat{\psi}_{|g\rangle}(z) \nonumber \\
    &+& \int dz \, \hat{\psi}^{\dagger}_{|e\rangle}(z)
    e^{-i k_p z} e^{-i \phi(t)} \hat{\psi}_{|g\rangle}(z) \bigg]  + \rm{h. \,
    c. \,},  \nonumber \\
    \end{eqnarray}
where h. c. is the Hermitian conjugate. The coupling strength $g =
d \sqrt{\omega_a/(2 \hbar \epsilon_o V)}$ between atoms and
photons in the pump light \citep{Fallani_PRA_2005} depends on
induced electric dipole moment $d$ of the atom, volume $V$ of the
condensate interacting with pump beam and permittivity
$\epsilon_o$. The phase difference between the pump and the
scattered photons is
    \begin{eqnarray}
    \label{eq:Phase}
    \phi(t) = \delta (1 - \epsilon \sin{\Omega t}) t,
    \end{eqnarray}
where $\delta = \omega_p - \omega_s$ is their frequency
difference. Frequency of the pump laser modulates at $\epsilon
\sin{\Omega t}$ which evolves $\phi(t)$ during the atom-photon
interaction time. The perturbing modulation amplitude is
small, $\epsilon \delta < \delta$.
Modulation of $\omega_p$ can be obtained by coupling
the pump light to an unstable Febry-Perot cavity. One mirror of the cavity is firmly
fixed and the external modulation is coupled to the other mirror, which
changes the cavity length as shown in Fig.
\ref{Fig:Experiment} (a). A small fraction of purely linear polarized
pump light can be coupled to the cavity
through a combination of polarizing beam splitter (PBS) and
$\lambda/4$-wave plate. The transmitted light leaked out of the
cavity will have $\pi/2$-rotated linear polarization after passing twice
through the $\lambda/4$-wave plate and hence reflects off from the
PBS. This reflected light can be detected by a fast photodiode (PD)
whose response time needs to be greater $(2\pi/\Omega)^{-1}$.
The PD detected signal can be used for frequency stabilization
of the pump laser by Pound-Drever-Hall locking technique
\citep{Drever_APB_1983}. The moving mirror of the cavity will shift
the lock point of $\omega_p$, which then will be modulated.
After releasing from ODT. the expanded BEC will fall through a
large cross section pump light after a delay of 1-2 ms.
The atom-photon interaction will be controlled by pulsing
the pump light. After some time of flight the BEC will be imaged for
quantitative measurement of the population in the ground state.
Since the pump light is approximately $- 2$ GHz detuned from
$\omega_a$ (Fig. \ref{Fig:Experiment} c), the excited state
population can be eliminated adiabatically using the Heisenberg equation of motion
$\dot{\hat{\psi}}_{|e \rangle} = i /\hbar \,
[\hat{H},\hat{\psi}_{|e\rangle}]$. In this case the ground state
population is
    \begin{eqnarray}
    \label{eq:BECPopulation}
    N = \int dz \, \hat{\psi}_{|g\rangle}^{\dagger}(z)
    \hat{\psi}_{|g\rangle}(z),
    \end{eqnarray}
which is the total number of atoms in the condensate.
This yields an equation of motion for the atoms
    \begin{eqnarray}
    \label{eq:GroundstateEquationOfMotion}
    \frac{d}{dt}\hat{\psi}_{|g\rangle}(z) &=&
    i \frac{\hbar}{2m}\nabla^{2} \hat{\psi}_{|g\rangle}(z) - i \frac{2
    g^{2}}{\Delta}\hat{\widetilde{c}}_p^{\dagger}
    \hat{\widetilde{c}}_p \nonumber \\
    &~&\bigg [1+\cos{\left(2k_p z + \phi(t) \right)} \bigg ]
    \hat{\psi}_{|g\rangle}(z),
    \end{eqnarray}
where $\Delta = \omega_{p}-\omega_{a}$ is the detuning of the pump
laser light and $\hat{\widetilde{c}}_p = \hat{c}_p e^{i
\omega_{p}t}$. Photons in the pump laser beam acquire an
equation of motion
    \begin{eqnarray}
    \label{eq:PhotonEquationOfMotion} \frac{d}{dt}
    \hat{\widetilde{c}}_p &=& - i \frac{2 g^{2}}{\Delta}
    \hat{\widetilde{c}}_p \int dz \,
    \hat{\psi}_{|g\rangle}^{\dagger}(z) \cos \left( 2k_p z + \phi(t)
    \right) \hat{\psi}_{|g\rangle}(z) \nonumber \\
    &~& -\kappa \hat{\widetilde{c}}_p + \eta,
    \end{eqnarray}
where $\kappa \leq c/2L $ is the damping rate of photons in the
BEC of length $L$ and $c$ is the velocity of light. Now in the following,
the bosonic operators $\hat{\psi}_{|g\rangle}$ and
$\hat{\widetilde{c}}_p$ are substituted by the coherent condensate wavefunction
$<\hat{\psi}_{|g\rangle}> = \psi_{|g\rangle}$ and the classical
light field amplitude $\widetilde{c}_p$ respectively.
Since $L$ is much longer than the radiation wavelength, one can
apply periodic boundary condition. We also assume homogeneous
density distribution for simplifying the calculation. In that case
evolution of the ground state wave-function can be written in the
basis of eigenfunctions
    \begin{eqnarray}
    \label{eq:GroundstateWavefunction}
    \psi_{|g\rangle}(z,t) = \sum_{n=0}^\infty  U_{n}(t) \, e^{2i n k_p z} e^{i n \phi(t)},
    \end{eqnarray}
where eigenvalues are $n \cdot \Delta p$ for $n = 0,1,2 \ldots$
and population at the $n^{\rm{th}}$ eigenstate is $\rho_n(t) =
U_{n}(t)^{\star} U_{n}(t)$. This atomic motion arises under the
assumption that the atoms in the BEC are delocalized
and their momentum uncertainty is negligible. Using
$\psi_{|g\rangle}(z,t)$ from Eq.
\eqref{eq:GroundstateWavefunction} in the Eqs.
\eqref{eq:GroundstateEquationOfMotion} and
\eqref{eq:PhotonEquationOfMotion}, three coupled ordinary
differential equations are obtained. Equation of motion of atoms
in the $n^{\rm{th}}$ level with momentum $p_0$ is
    \begin{eqnarray}
    \label{eq:ReducedGroundstateEquationOfMotion}
    \frac{d}{dt}U_{n} = &-& i4n^2\omega_{r} U_{n} - i n \dot{\phi}(t)
    U_{n} \nonumber \\
    &-& i \frac{g^{2}N}{\Delta} \alpha [2U_{n} + U_{n+1}].
    \end{eqnarray}
Similarly equation of motion of atoms shifted to the
$(n+1)^{\rm{th}}$ level with momentum $p_0 + \Delta p$ is
    \begin{eqnarray}
    \label{eq:ReducedShiftedGroundstateEquationOfMotion}
    \frac{d}{dt}U_{n+1} = &-& i4(n+1)^2\omega_{r} U_{n+1} - i (n+1) \dot{\phi}(t)
    U_{n+1} \nonumber \\
    &-& i \frac{g^{2}N}{\Delta} \alpha [2U_{n+1} + U_{n}].
    \end{eqnarray}
Equation of motion of photons in the pump laser beam is
    \begin{eqnarray}
    \label{eq:ReducedPhotonEquationOfMotion}
    \frac{d}{dt}\widetilde{c}_p = - i \frac{2g^{2}N}{\Delta}
    \widetilde{c}_p \rho_{n,n+1} -\kappa \widetilde{c}_p +
    \eta,
    \end{eqnarray}
where $\omega_r = \hbar k_p^2/2 m$ is the single photon recoil
frequency, $\alpha = \widetilde{c}_p^{\dagger} \widetilde{c}_p$
is the intensity of the light field and $\rho_{i,j} = U_{i}^{\star} U_{j}$ is the
coherence between $i^{\rm{th}}$ and $j^{\rm{th}}$ eigenstates. The
quantity $\rho_{i,j}$ reduces to the density of a state when $i =
j$. We have introduced rescaled light amplitude
$\widetilde{c}_p \rightarrow\sqrt{N} \widetilde{c}_p$
in the Eq. \eqref{eq:ReducedPhotonEquationOfMotion}.
\begin{figure}
\begin{center}
\resizebox{0.4\textwidth}{!}{
\includegraphics{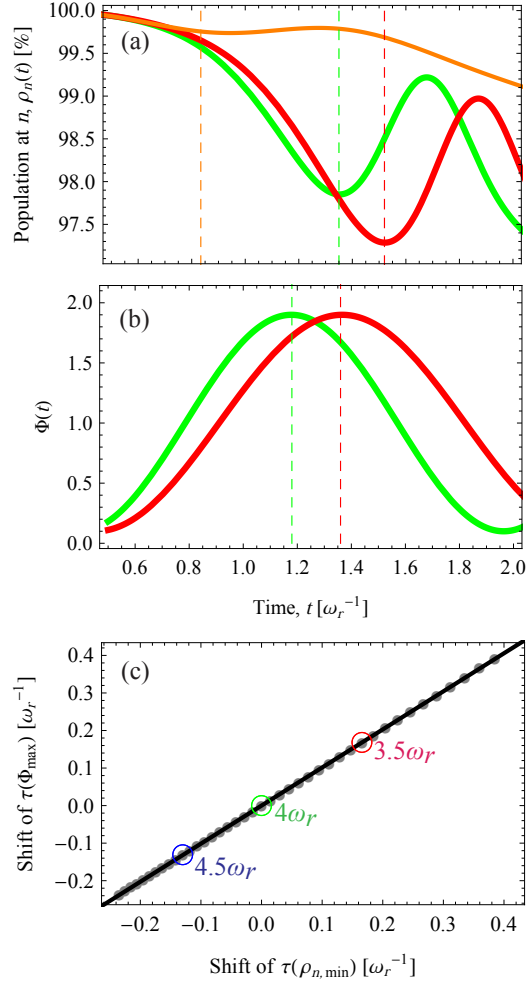}
}
\end{center}
\caption[Population and Phase]{(Color online) The solid lines show
(a) population at the initial momentum state, (b) $\Phi(t)$ at
modulation frequencies $\Omega = 4 \omega_r$ (green), $3.45
\omega_r$ (red) and $1.2 \omega_r$ (orange). A strong modulation
of the pump laser at $\epsilon = 0.8$ is considered for all
spectra. The dashed lines show $\tau_i$ where the first minima in
the $\rho_{n, \, \rm{min}}$ and the first maxima $\Phi_{\rm{max}}$
appears, which are distinguished by their respective colors. (c)
Shift of $\tau(\Phi_{\rm{max}})$ as $\tau(\rho_{n, \, \rm{min}})$
shifts with change of $\Omega$ from blue (blue) to red (red) detuning relative
to $4 \omega_r$ (green) and the solid line is a linear fit to it.}
\label{Fig:PopulationPhaseShift}
\end{figure}
For simplicity we consider the mass flow occurs from $n^{\rm{th}}$ to
a $(n+1)^{\rm{th}}$ level. That gives a population difference between two
momentum states
    \begin{eqnarray}
    \label{eq:PopulationDifference}
    \Delta \rho (t) &=& \rho_{n}(t) - \rho_{n+1}(t),
    \end{eqnarray}
and $\rho_{n} + \rho_{n+1} = N/V$ remains conserved. A new set of
equations can be obtained for the rate of change of coherence
function
    \begin{eqnarray}
    \label{eq:CorrelationFunction}
    \frac{d}{dt}{\rho}_{n,n+1} &=& i[4(1+2n) \omega_{r}
    + \dot{\phi}(t)] \rho_{n,n+1} \nonumber \\
    &~& - \mathcal{D} \, \rho_{n,n+1}
    + i \, \alpha \frac{g^{2}N}{\Delta} \Delta \rho,
    \end{eqnarray}
and for the population difference
    \begin{eqnarray}
    \label{eq:PopulationDifference}
    \frac{d}{dt}\Delta \rho = - 4 \alpha \frac{g^{2}N}{\Delta}
    \Im \left[ \rho_{n,n+1} \right],
    \end{eqnarray}
where, $\Im$ indicates imaginary part of the correlation function.
The decoherence rate of atoms $\mathcal{D}$ arises due to Doppler
broadening, inhomogeneity in the condensate and phase diffusion.
The phase diffusion mechanism depends on $\delta$ and $p_o$.
Equations
\eqref{eq:CorrelationFunction}-\eqref{eq:PopulationDifference},
are then equivalent to the well known Maxwell-Bloch
equations for a two-level system
\citep{Inouye_PRL_2000,Piovella_OC_2001}.
\begin{figure}
\begin{center}
\resizebox{0.4\textwidth}{!}{
\includegraphics{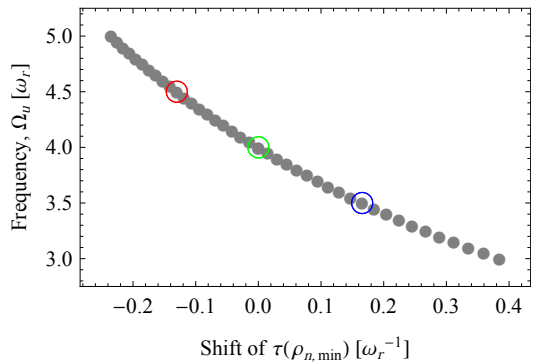}
}
\end{center}
\caption[Population and Unknown frequency]{At $\Omega_r=4\omega_r$
(green) the relation between $\Omega_u$ and shift of
$\tau(\rho_{n, \, \rm{min}})$. Example blue and red detuned
$\Omega_u$ are indicated by their colors.}
\label{Fig:UnknownFrequency}
\end{figure}
Solution of the Eqns. \eqref{eq:CorrelationFunction} and
\eqref{eq:PopulationDifference} identify quantum and semiclassical
regimes of the superradiance scattering \citep{Fallani_PRA_2005}.
In the quantum regime, $g^2 \sqrt{N}/ \omega_r \Delta < 2
\sqrt{\kappa/ \omega_r}$, momentum of each atom changes by $\Delta
p$ due to scattering of single photon at small $\kappa$.
At larger $\kappa$, multiple scattering of photons give rise to
momentum change of atoms greater than $\Delta p/2$. This
semiclassical regime starts at $g^2 \sqrt{N}/ \omega_r \Delta > 2
\sqrt{\kappa/ \omega_r}$. We focus on quantum superradiance for
detection of a frequency. The CARL phenomena starts when the
build up laser power $\alpha(t) = \widetilde{c}_p^\dagger \widetilde{c}_p$
in the condensate reaches the threshold, which results in the
population transfer to $(n+1)^{\rm{th}}$-level. Coherence time
of the matter wave grating determines the efficiency
of the superradiant process. The matter wave grating has a finite
lifetime which depends on $p_0$ and the superradiance fades off
with the decay of the matter wave grating.
Detection of an unknown frequency $\Omega_u$ requires measurement
of the time evolution of $\rho_n(t)$ as shown in Fig.
\ref{Fig:PopulationPhaseShift} (a), when the pump laser frequency
is modulated at $\Omega_u$. A measurement of the
$\rho_n(t)$-spectrum at a known modulation frequency $\Omega_r$
while keeping rest of the conditions unchanged is required. Antinodes in these
spectra corresponds to the maximum population transfer to
the $(n+1)^{\rm{th}}$ level. The time $\tau_i(\rho_{n, \rm{min}})$
corresponding to the first antinode appears at later times for a
higher $\Omega_i$, where the subscript $i$ refers to the different
modulation frequencies. For our interest, measurement of the
$n^{\rm{th}}$ level population at least up to the time
$\tau_i(\rho_{n, \rm{min}})$ is necessary. The phase modulating
part  $\Phi(\Omega, \, t) = \left( 1 - \epsilon \sin{\Omega t}
\right)$, from Eq. \eqref{eq:Phase}, evolves in time as shown in
Fig. \ref{Fig:PopulationPhaseShift} (b) for two different
$\Omega_i$. Like $\tau_i(\rho_{n, \rm{min}})$, the time
$\tau_i(\Phi_{\rm{max}})$ at which the first antinode appears,
also shifts by the same amount with $\Omega_i$. Figure
\ref{Fig:PopulationPhaseShift} (c) shows one-to-one shifts of
$\tau_i(\rho_{n, \rm{min}})$ and $\tau_i(\Phi_{\rm{max}})$
with the change of $\Omega_i$ with respect to $\Omega_r = 4
\omega_r$.
For frequency detection, we first need to determine
$\tau_r(\rho_{n, \rm{min}})$ and $\tau_u(\rho_{n, \rm{min}})$ from
experimentally measured $\rho_n(t)$-spectra at known and unknown
modulations $\Omega_r$ and $\Omega_u$ respectively. For
modulations at $\Omega_r$, the condition of minima of the
$\rho_n(t)$-spectra
    \begin{eqnarray}
    \label{eq:Minima}
    \frac{d\Phi(\Omega_i, \, t)}{dt} = 0,
    \end{eqnarray}
determines $\tau_r(\Phi_{\rm{max}}) = \pi/2\Omega_r$ for the first
antinode. With this, $\tau_u(\Phi_{\rm{max}})$ for the unknown
frequency is
    \begin{eqnarray}
    \label{eq:TauR}
    \tau_u(\Phi_{\rm{max}}) = \tau_r(\Phi_{\rm{max}}) \pm
    |\tau_u(\rho_{n, \rm{min}})-\tau_r(\rho_{n, \rm{min}})|,
    \end{eqnarray}
where $\tau_i(\rho_{n, \rm{min}})$ are measured experimentally. In
this equation $``+"$ appears when $\tau_u(\rho_{n, \rm{min}}) >
\tau_r(\rho_{n, \rm{min}})$ and $``-"$ when the reverse condition
is true. Equation \eqref{eq:Minima} determines the unknown
frequency $\Omega_u = \pi/2 \tau_u(\Phi_{\rm{max}})$ using
$\tau_u(\Phi_{\rm{max}})$ from Eq. \eqref{eq:TauR}. Figure
\ref{Fig:UnknownFrequency} shows sample detection of $\Omega_u$
with respect to the shifts of $\tau(\rho_{n,
\rm{min}})$ estimated from measured time evaluation of
$\rho_n(t)$. The rage of $\Omega_u$ which can be detected depends on
fluctuation limited experimental resolution for measuring the
change of population in the $n^{\rm{th}}$ level. As for example
Fig. \ref{Fig:PopulationPhaseShift} (a) shows a
$\rho_n(t)$-spectrum at $\Omega = 1.2 \omega_r$ from which the
$\tau(\rho_{n, \rm{min}})$ will be difficult to extract due to
statistical fluctuations in the measurement. Assuming an
experiment can resolve $\sim 1\%$ atom number fluctuation, a range
of frequencies from $2.5 \omega_r - 6 \omega_r$ would be possible
to detect by this method. As for example recoil frequency $\omega_r =
2\pi \times 3.7$ kHz and $2\pi \times 63$ kHz for the
D2-transition of $^{87}\rm{Rb}$ and $^7\rm{Li}$ isotopes
respectively. This estimates a range of frequencies from few
kHz to about a MHz could be detected by this technique.
Resolutions of the measured $\tau_i(\rho_{n, \rm{min}})$ could be
accurate to $\sim 2$ ns using precise pulse train generators
\citep{PulseBlaster} as a timing system. This gives a frequency
resolution $\Delta \Omega = s \omega_r^2/\pi \times 10^{-9}$,
where $s \simeq 10/3$ is approximate slope of the spectrum shown
in Fig. \ref{Fig:UnknownFrequency}. As for example $\Delta \Omega
= 2\pi \times 0.1$ Hz for $^{87}\rm{Rb}$ and which gives a
fractional uncertainty $\sim 10^{-6}$. A state-of-the-art
frequency standard requires flywheel oscillators
\citep{Allan_IEEE_1975}, since the atomic frequency standards have
limited run time. The proposed techniques could be incorporated
for frequency stabilization of the flywheel oscillators.
The authors thank Amitava Sen Gupta for useful discussions. A.
Bhattacherjee acknowledges financial support from the Department
of Science and Technology, India (grant SR/S2/LOP-0034/2010) and
S. De carried out this work as a part of the STIOS program funded
by CSIR-NPL.

\end{document}